\documentclass[nofootinbib,a4paper,fleqn,aps,prd,10pt,superscriptaddress,reprint,showkeys,showpacs]{revtex4-1}

\usepackage{times}
\usepackage{amsfonts}
\usepackage{amsmath}
\usepackage{amssymb}
\usepackage{natbib}
\usepackage{graphicx}
\usepackage{enumerate}
\usepackage[inline]{enumitem}
\usepackage{paralist}
\usepackage{xcolor}
\usepackage[colorlinks=true,linkcolor=blue]{hyperref}
\usepackage[outdir=./]{epstopdf}
\usepackage[normalem]{ulem}

\def\aap{A\&A}
\def\aj{AJ}
\def\apj{ApJ}
\def\apjl{ApJL}
\def\apjs{ApJS}
\def\apss{Astrophysics and Space Science}
\def\jcap{JCAP}
\def\mnras{MNRAS}

\def\physrep{Physics Reports}

\newcommand{\Mov}[1]{{\color{black}{#1}}}

\begin{document}

\title{{\bf Non self-similar Luminosity-temperature relation and dynamical friction
}}

\author{Antonino~\surname{Del Popolo}}%
\affiliation{%
Dipartimento di Fisica e Astronomia, University of Catania, Viale Andrea Doria 6, 95125, Catania, Italy
}
\affiliation{%
INFN sezione di Catania, Via S. Sofia 64, I-95123 Catania, Italy
}
\affiliation{Institute of Astronomy, Russian Academy of Sciences, 119017, Pyatnitskaya str., 48 , Moscow}
\email{adelpopolo@oact.inaf.it}

\author{Morgan~\surname{Le~Delliou}}
\email[Corresponding author: ]{(delliou@lzu.edu.cn,) Morgan.LeDelliou.ift@gmail.com}
\affiliation{Institute of Theoretical Physics, School of Physical Science and Technology, Lanzhou University, No.222, South Tianshui Road, Lanzhou, Gansu 730000, China}
\affiliation{Instituto de Astrof\'isica e Ci\^encias do Espa\c co, Universidade de Lisboa, Faculdade de Ci\^encias, Ed. C8, Campo Grande, 1769-016 Lisboa, Portugal}
\affiliation{Lanzhou Center for Theoretical Physics, Key Laboratory of Theoretical Physics of Gansu Province, Lanzhou University, Lanzhou, Gansu 730000, China}
 \author{Man~Ho~\surname{Chan}}
 \email{chanmh@eduhk.hk}
 \affiliation{Department of Science and Environmental Studies, The Education University of Hong Kong, Tai Po, New Territories, Hong Kong}
%
%

\label{firstpage}

\date{\today}

\begin{abstract}
Extending the results of a previous paper \citep{DelPopolo2005}, by taking into account the role of dynamical friction, we 
recovered 
the luminosity-temperature relation 
(LTR)
. While 
by 
assuming self-similarity\Mov{,} a scaling law in which $L\propto T^2$ is obtained, observations show that the relation between luminosity and temperature is steeper, 
${L \propto T^ {\simeq 3}}$.
This difference can be explained in terms of energy input by non-gravitational processes, like pre-heating, supernovae feedback, and heating from AGN. 
In this paper, we studied the LTR by means of a modified version of the punctuated equilibria model \citep{Cavaliere1999}, taking into account 
in addition 
dynamical friction, 
thus extending the approach 
\Mov{found }in 
\citep{DelPopolo2005}. The result is a non-self-similar LTR with a bend at $\simeq 2$ keV, with a slope $2.76 \pm 0.18$ at larger energies and $3.4 \pm 0.18$ at energies smaller than 2 keV.
This result is in agreement with the XXL survey \citep{Giles2016}. Moreover the steeper slopes at smaller energies is in agreement with some studies claiming a further steepening of the LTR at the low mass end. We also compared the results of our model with the 400d groups sample, finding that in groups the slope is slightly steeper than in clusters, 
namely 
$3.35 \pm 0.3$, in agreement with 
the 
\citep{Zou2016} study for the 400d groups sample, 
that gives 
a slope $3.29 \pm 0.33$. 
\end{abstract}

\pacs{98.52.Wz, 98.65.Cw}

\keywords{Dwarf galaxies; galaxy clusters; modified gravity; mass-temperature relation}

\maketitle

\section{Introduction}

Galaxy clusters represent the largest gravitationally bound systems in the Universe, with dimensions that can vary in the range $2-10$ Mpc, constituted by thousands or tens of thousands of galaxies, emitting a large amount of energy in the X band with powers in the range $10^{43}-10^{45} \rm erg/s$, coming from low-density intra-clusters medium (ICM) ($\simeq 12 \% $ of the total mass). Apart 
from the 
ICM, galaxy clusters are dominated by dark matter (DM) ($\simeq 85 \% $ of the total mass), and also contain stars ($\simeq 3 \% $ of the total mass). There are several reasons why we are interested in the study of clusters. Three of these are 
\begin{inparaenum}[\itshape a\upshape)]
 \item the formation and evolution of clusters itself, 
 \item the use of clusters as probe in cosmology, and 
 \item the study of   
extreme physical processes, 
such 
as \begin{inparaenum}[\itshape i\upshape)]
\item the interaction between jets in active galactic nuclei (AGN), 
\item the plasma of the ICM emitting in the X-ray band, or 
\item cluster mergers, also used to have information on the characteristics of the DM, as in the case of the Bullet cluster \citep{Clowe2006}.
\end{inparaenum}
\end{inparaenum}
  Past observations (e.g., ROSAT, ASCA), have shown the existence of tight correlation between fundamental parameters in clusters like the total cluster mass, $M_{\rm tot}$, their X-ray luminosity, $L_{\rm X}$, and the temperature of the ICM \citep{Markevitch1998,Horner1999}. The relationships between those quantities can \Mov{often }be 
  expressed in terms of power laws, like the mass-temperature relation (MTR), the luminosity-temperature relation (LTR), and can be used to have ``cheap" mass estimates for a noteworthy number of clusters, to high redshifts \citep{Maughan2007}. On the other hand, X-ray temperature gives information on the depth of the cluster potential well, and the bolometric luminosity, $L \propto n^2 R_{\rm X}^3 T^{1/2}$,
emitted as bremsstrahlung by the ICM, 
as well as 
on the baryon number density, $n$, in a given volume proportional to the radius 
cubed 
$R_{\rm X}^3$. Scaling relations 
with power-law shape 
are expected 
from 
the simplified assumption that clusters are self-similar, 
and 
formed in a sort of monolithic collapse, 
the ICM being 
heated by the shocks generated 
from 
the collapse. This simple model was introduced many years ago by Kaiser \citep{Kaiser1986}. In the self-similar model clusters of galaxies, 
of any sizes
, are identical when scaled by their mass. In this case one speaks of ``strong self-similarity" \citep{Bower1997}, related to the redshift independence of the power law slopes. 
However, 
a redshift dependent evolution of the scaling relations' normalization 
is expected
, dubbed ``weak self-similarity". The $L_{\rm X}-kT$ (LTR) relation is the 
most studied \citep{Mushotzky1984,Ettori2004,Mittal2011}, because these two properties can be directly measured 
from X-ray data\Mov{, and moreover each can be measured independently }\citep{Maughan2012}. 
Kaiser's model \Mov{predicted }
a dependence between $L$, and $T$, as $L \propto T^2$, while a large number of studies showed that $L \propto T^3$ \citep{Pratt2009,Connor2014}
, \Mov{observing a steeper }
slope 
than the self-similar prediction. 
This 
implies that low mass clusters are hotter and/or less luminous than what the self-similar models predicts. We should stress that when the core regions of the most massive, relaxed clusters are not considered, they show self similar behavior \citep{Maughan2012}. One of the interpretations of this effect is that in low mass haloes, having a weaker gravitational potential, non-gravitational heating has a stronger impact on the ICM, 
leading to the breaking of self-similarity. There are several explanations for the deviation from the self-similar, $L \propto T^2$ law, and several possibilities to obtain a scaling law close to the one observed. 
One possibility is to 
inject 
the 
non-gravitational energy into the ICM, during the period of formation of clusters. 
This solution is usually 
called pre-heating. 
%
%
Other effects are 
supernovae-feedback heating from AGN at high redshift. 
Such 
processes have a larger effect on lower mass systems, since 
the latter 
are characterized by shallower potential wells, expelling \Mov{more }gas from the inner regions with the consequence of reducing the luminosity. 

Observations of the clusters cores \Mov{give }
other evidences for the existence of non-gravitational processes/heating, both in groups and in clusters. In fact in the core regions, the ICM density is high
, which 
leads to 
shorter 
cooling times 
than the cluster's lifetime. As a consequence a cooling flow should be generated. In the cooling flow, cooling 
gas condenses 
in the core of the cluster and is replaced by inflow of gas from larger radii
, that in turn cools and therefore subsequently flows 
towards the core. However, 
the 
high cooling rate predicted by this model has not been observed. Energy input from AGN could be the mechanism to balance cooling in the core of 
clusters. 
Such 
AGN input could break the self-similarity in clusters, and 
balance 
the cooling in the cores. Self-similarity 
can thus be 
broken by a sort of heating from AGN producing an increase of the gas entropy in the \Mov{cluster}
, while reducing its density. The study of the $L-T$ relation in groups, and small clusters can help to understand the nature of the non-gravitational processes which \Mov{break }
the self-similarity. Unfortunately, measuring the $L-T$ relation in groups is more complicated than in clusters, because they are less luminous. Nevertheless, some studies have \Mov{reached }
a sort of consensus\Mov{, namely} that the $L-T$ relation is similar \Mov{to that found in }
\citep{Osmond2004} or steeper than \citep{Helsdon2000} 
in clusters. 

X-ray flux limited sample are subject to two forms of selection bias. One is the Malmquist bias, in which clusters having higher luminosity are preferentially selected out to higher redshift. The other is the Eddington bias. In this case objects above a flux limit will show average luminosity for their mass. This happens in presence of statistical or intrinsic scatter in luminosity for a given mass. \Mov{Their} effect 
on 
the $L-T$ relation is 
to bias both normalization and slope
. It is consequently 
of fundamental importance to take account of the bias when one is modeling the scaling relations in clusters, to understand the nature of the non-gravitational heating breaking the self-similarity. Selection bias can explain the departure from self-similarity \citep{Maughan2012}.

As already reported, some studies \citep[e.g.][]{Helsdon2000,Sun2009} claim a further steepening of the LTR at the low mass end. However, 
recent works, correcting for sample selection effect, have suggested that in reality the LT relation slope in groups and in clusters are consistent \citep{Lovisari2015}. In the present paper, we use a modification of the punctuated equilibria model (MPEM) of \citet{Cavaliere1999}. 
Ref.~\citep{DelPopolo2005} already 
presented a model for the LTR based on an extension of the punctuated equilibria model, where the luminosity depended from the angular momentum, and the cosmological constant. In the present paper, we add the effect of dynamical friction 
(DF)
, and compare the LTR with the results from the XXL survey \citep{Giles2016}, and 
with those 
of \citep{Zou2016} for galaxy groups. As predicted by some authors \citep{Helsdon2000,Sun2009}, going to the low mass end, and 
to 
galaxy groups, a slight steepening of the LTR relation is visible.

The paper is organized as follows. In Sec.~\ref{sect:model}, we discuss the mass-temperature relation. In Sec.~\ref{sec:pem}, we show how the LTR is obtained through an improved punctuated equilibria model. Sec.~\ref{sect:results} is devoted to results and discussion. Finally, Sec.~\ref{sect:conclusions} \Mov{presents our }
conclusions. 

%
%
\section{The mass-temperature relation}\label{sect:model}

In the calculation of the \Mov{LTR}
, we will use the \Mov{MTR}
, a relation between the X-ray mass of clusters and their 
temperature, $T_{\rm X}$. The MTR can be obtained by means of 
scaling arguments. The mass \Mov{with}in the virial radius is given by $M(\Delta_{\rm vir}) \propto T_{\rm X}^{3/2} \rho_{\rm c}^{-1/2} \Delta_{\rm vir}^{-1/2}$, where $\rho_{\rm c}$ is the 
critical density, and $\Delta_{\rm vir}$ the density contrast of a spherical top-hat perturbation after collapse and virialization. 

This result obtained by simple scaling arguments can be obtained in much more detail using two different methods: 
\begin{inparaenum}[\itshape a\upshape)]
 \item the so called ``late formation approximation" based on an improvement of the spherical collapse model.\\ In \cite{DelPopolo2002} the simple spherical collapse model was improved taking into account the effect of angular momentum, the cosmological constant, and a modified version of the virial theorem, including a surface pressure term \citep{Voit1998,Voit2000,Afshordi2002,DelPopolo2005}. In \cite[Sec.~III.A]{DelPopolo2019} 
 has 
 in addition 
 been taken 
 into account the effect of 
 DF
 . 
The ``late-formation approximation" is based on the idea that clusters form from a top-hat density profile. It also assumes that the redshift of observation, $z_{\rm obs}$, is equal to 
that of formation, $z_{\rm f}$. 
Such approximation presents 
pros and cons. 
It is 
valid 
in the case $\Omega_{\rm m,0} \simeq 1$, where 
cluster formation is fast, and at all redshift $z_{\rm obs} \simeq z_{\rm f}$. For a different value of $\Omega_{\rm m,0}$ it is necessary to take into account the difference between $z_{\rm obs}$ and $z_{\rm f}$. 
Another problem is that, as shown by \citep{Voit2000}, to get the correct normalization of the MTR and its time evolution, continuous accretion is needed .\\
\item 
\Mov{the }
top-hat model (spherical collapse model)\Mov{, in which structures form continuously,} is substituted by a model of cluster formation from spherically symmetric perturbations with negative radial density gradients. The gradual way clusters form is obtained by means of the merging-halo formalism of \citet{Lacey1993}. 
\end{inparaenum}

In the following, we specialize to the second model, also discussed in \cite[Sec.~III.B]{DelPopolo2019}.

Here we summarize the calculation for 
the convenience of 
the reader.   
 
In the following, we consider some gravitationally growing mass concentration collecting into a potential well. Let ${\rm d}P=f(L,r,v_r,t){\rm d}L{\rm d}v_r{\rm d}r$ be the probability that a particle, having angular momentum $L=r v_{\theta}$, is located 
within 
$[r,r+{\rm d}r]$, with velocity ($v_r={\dot r}$) 
in 
$[v_r,v_r+{\rm d}v_r]$, and angular momentum 
in 
$[L,L+{\rm d}L]$. 
The angular momentum that we consider takes into account the ordered angular momentum (namely generated by tidal torques) and random angular momentum \citep[see][Appendix~C.2]{DelPopolo2009}. 
The radial acceleration of the 
particle is: 
\begin{equation}\label{eq:coll}
 \frac{{\rm d}v_r}{{\rm d}t} = -\frac{GM}{r^2} + \frac{L^2(r)}{M^{2}r^3} + \frac{\Lambda}{3}r -
 \eta\frac{{\rm d}r}{{\rm d}t}\,,
\end{equation}
\citep[see][]{Peebles1993,Bartlett1993,Lahav1991,DelPopolo1998,DelPopolo1999}
with $\Lambda$ being the cosmological constant and $\eta$ the 
DF 
coefficient. 
The last term, the 
DF 
force per unit mass, is given in \citep[Appendix~D, Eq.~D5]{DelPopolo2009}. Note that the previous equation can be obtained via Liouville's theorem \citep{DelPopolo1999}.

Integrating Eq.~(\ref{eq:coll}) with respect to $r$ we have:
\begin{equation}\label{eq:coll1}
 \frac{1}{2}\left(\frac{{\rm d}r}{{\rm d}t}\right)^{2} = \frac{GM}{r} + 
 \int_0^r \frac{L^{2}}{M^{2}r^{3}}\mathrm{d}r + \frac{\Lambda}{6}r^{2} - \int_0^r 
 \eta\frac{{\rm d}r}{{\rm d}t} + \epsilon\,.
\end{equation}
The specific binding energy of the shell, $\epsilon$, can be obtained from the turn-around condition 
$\frac{{\rm d}r}{{\rm d}t}=0$.

Integrating again Eq.~(\ref{eq:coll1}), one gets:
\begin{equation}\label{eq:tmppp}
 t = \int\frac{\mathrm{d}r}{\sqrt{2\left[\epsilon + \frac{GM}r+\int_{r_{\rm i}}^{r} 
     \frac{L^2}{M^2r^3}\mathrm{d}r+\frac{\Lambda}{6}r^2\right] -\int \eta\frac{{\rm d}r}{{\rm d}t}{\rm d}r}}\,.
\end{equation}

The specific energy of infalling matter, following \citep{Voit2000}, can be written as 
\begin{equation}
 \epsilon_l = -\frac{1}{2}\left(\frac{2\pi GM}{t_\Omega}\right)^{2/3}
               \left[\left(\frac{M_0}{M}\right)^{5/(3m)}-1\right]g(M)\,,
\end{equation}
where $m=5/(n+3)$ is a constant specifying how the mass variance evolves as a function of $M$, 
$n$ is the usual power-law perturbation index in wavenumber space, $M_0$ is a fiducial mass,  $t_{\Omega}=\pi\Omega_{\rm m,0}/[H_0(1-\Omega_{\rm m,0})^{3/2}]$.
The function $g(M)$ is given by
\begin{equation}
 g(M) = 1+\frac{F}{x-1} +\frac{\lambda_0}{1-\mu(\delta)}+\frac{\Lambda}{3 H^2_0 \Omega_{\rm m,0}} \xi^3\,,
\end{equation}
where $x=1+(t_{\Omega}/t)^{2/3}$, $M=M_0 x^{-{3m/5}}$ \cite{Voit2000}, $M_0$ is given in \citep{Voit2000}, $\xi=r_{\rm ta}/x_1$, where $r_{\rm ta}$ is the turn-around radius, and $x_1$ is defined by $M=4 \pi \rho_b x_1^3/3$, where $\rho_b$ is the background density
and
\begin{equation}
 F = \frac{2^{7/3}\pi^{2/3}\bar{\rho}_{\rm m,0}^{2/3}}{3^{2/3}H^2_0\Omega_{\rm m,0}M^{8/3}}
     \int_{r_{\rm i}}^{r_{r_{\rm ta}}} \frac{L^2}{r^3}\mathrm{d}r\,.
\end{equation}

Integrating $\epsilon_l$ with respect to the mass \citep{Voit2000} we 
get $-\int \epsilon_l dM=E/M$, where $E$ is the kinetic energy. Then we can connect temperature and the ratio $E/M$ with the equation
\begin{equation}
 k_{\rm B}T = \frac{4}{3}\tilde{a}\frac{\mu m_{\rm p}}{2\beta}\frac{E}{M}\,,
\end{equation}
where $\tilde{a} = \frac{\bar{\rho}_{\rm m,vir}}{2\rho(r_{\rm vir}) - \bar{\rho}_{\rm m,vir}}$ is the ratio between 
kinetic and total energy \citep{Voit2000}, and $\bar{\rho}_{\rm m,vir}$ the mean density within the virial radius. 
After having calculated $E/M$, we obtain the energy given by
\begin{eqnarray}\label{eq:kT}
\frac{k_{\rm B}T}{\rm keV} & = & \frac{2}{5}\tilde{a}\frac{\mu m_{\rm p}}{2\beta} 
                                 \frac{m}{m-1}\left(\frac{2\pi G}{t_\Omega}\right)^{\frac{2}{3}}M^{\frac{2}{3}}
                                 \times\nonumber\\
                           & &   \left[\frac{1}{m} + \left(\frac{t_\Omega}{t}\right)^{\frac{2}{3}} + 
                                 \frac{K(m,x)}{(M/M_0)^{8/3}} + \frac{\lambda_0}{1-\mu(\delta)} \right.\nonumber\\
                           & &   \left.\quad +\frac{\Lambda\xi^3}{3H^2_0\Omega_{\rm m,0}}
\right]\,,
\end{eqnarray}
where $\lambda_0=\epsilon_0 T_{\rm c0}$, is given by  \cite[Eqs.~(23-24)]{AntonuccioDelogu1994}, and $\mu(\delta)$ is given in \cite[bottom of Eq.~(29)]{AntonuccioDelogu1994}.  
The function $K$ of $m$, and $x$, is expressed in terms of the ${\rm LerchPhi}$ function\footnote{\begin{equation}
 {\rm LerchPhi}(z,a,v) = \sum_{n=0}^{\infty}\frac{z^n}{(v+n)^a}\,.
\end{equation}, definition valid for $|z| < 1$. By analytic 
continuation, it is extended to the whole complex $z$-plane for each value of $a$.
}

\begin{eqnarray}
 K(m,x) & = & (m-1)F x {\rm LerchPhi}(x,1,3m/5+1) - \nonumber \\
        & & (m-1) F {\rm LerchPhi}(x,1,3m/5)\,,
\end{eqnarray}

Eq.~(\ref{eq:kT}) can be written as
\begin{equation}\label{eq:kT1}
 k_{\rm B}T \simeq 8~{\rm keV} \left(\frac{M}{10^{15}h^{-1}M_{\odot}}\right)^{\frac{2}{3}}\frac{m(M)}{n(M)}\,.
\end{equation}
where the normalization was obtained following \citep{Voit2000}.

The functions $m(M)$ and $n(M)$ are given by
\begin{eqnarray}
 m(M) & = & \frac{1}{m} + \left(\frac{t_\Omega}t\right)^{\frac{2}{3}} + \frac{K(m,x)}{(M/M_0)^{8/3}} + \nonumber\\
      &   & \frac{\lambda_0}{1-\mu(\delta)} + \frac{\Lambda\xi^3}{3H^2_0\Omega_{\rm m,0}} \,,\\
 n(M) & = & \frac{1}{m} + \left(\frac{t_{\Omega}}{t_{0}}\right)^{\frac{2}{3}} + K_0(m,x)\,,
\end{eqnarray}
where $K_0(m,x)$ indicates that $K(m,x)$ must be calculated assuming $t=t_0$.

Comparing to Eq.~(17) of \citep{Voit2000}, Eq.~(\ref{eq:kT1}) shows an additional mass-dependent term. As a result, the MTR shows a break at the low mass end, and is no longer self-similar.

In addition to the works of \citep{Voit1998,Voit2000}, an MTR and its scatter were also found by 
\citep{Afshordi2002}
. 
The results concerning the MTR and the scatter that we found here are in agreement with their results [see Eqs.~38, and 39 in \citealt{DelPopolo2019}, and the discussion in \citealt{Afshordi2002}].

%
%

\begin{figure*}[!ht]
 \centering
 \includegraphics[width=10cm,angle=0]{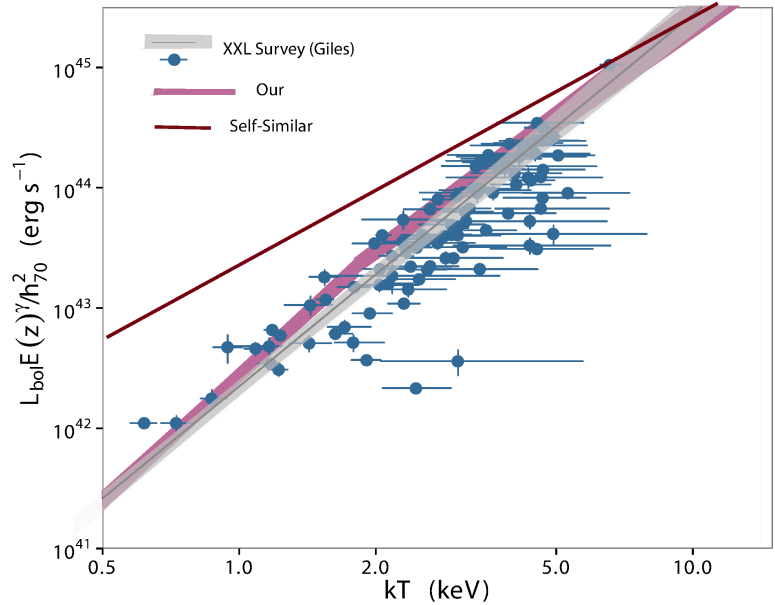}
 \caption[justified]{Comparison of the LTR \Mov{from }
 \citep{Giles2016}, represented by the dots with error-bars and the gray line, with our model. The plot shows the bolometric LTR, with data  corrected for the bias. The black solid line is the best fitting model, and the gray band represents the 1 $\sigma$ uncertainty. The red band represents our result, while the brown line is the prediction of the self-similar model, $L \propto T^2$. 
%
%
%
%
}
 \label{fig:comparison.1}
\end{figure*}

\begin{figure*}[!ht]
 \centering
 \includegraphics[width=10cm,angle=0]{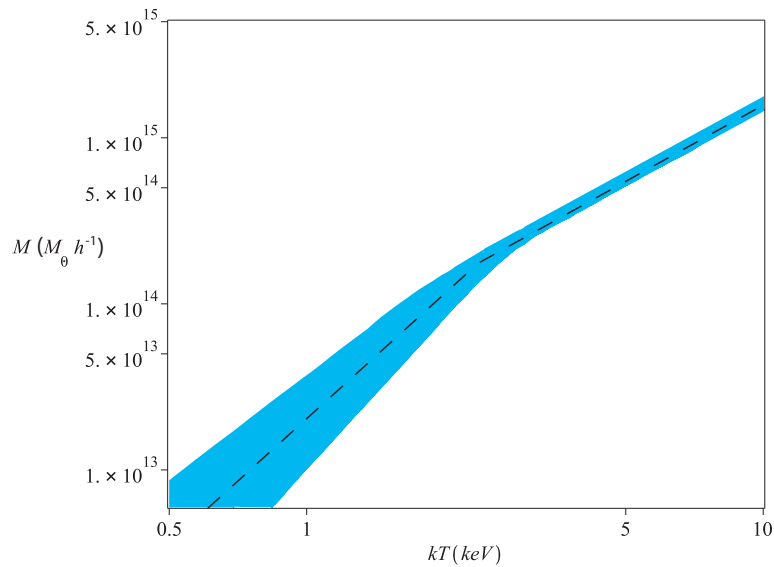}
 \caption[justified]{The MTR obtained with the model \Mov{from }
 Sec.~\ref{sect:model}. The dashed line is the average value. The cyan region shows the 68\% confidence level region, obtained according to \citep{Afshordi2002}, and \citep{DelPopolo2019}.
 }
 \label{fig:comparison.2}
\end{figure*}

\begin{figure*}[!ht]
 \centering
 \includegraphics[width=8cm,angle=0]{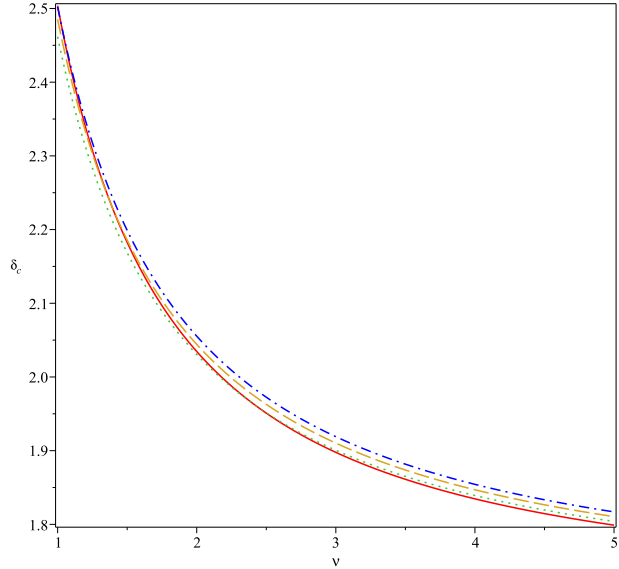}
 \caption[justified]{The collapse threshold $\delta_c(\nu)$ as a function of $\nu$. The red solid line shows the result \Mov{from }
 \citep[Eq.~4]{Sheth2001}, the green short-dashed line the results of \citep{DelPopolo1998}, taking into account the effect of the tidal field, the orange long-dashed the result of \citep{DelPopolo2006c} taking into account the effect of the tidal field and the cosmological constant, while the blue dot-dashed line takes into account the effect of the tidal field, the cosmological constant and 
 DF.}
 \label{fig:comparison.3}
\end{figure*}

\begin{figure*}[!ht]
 \centering
\includegraphics[width=10cm,angle=0]{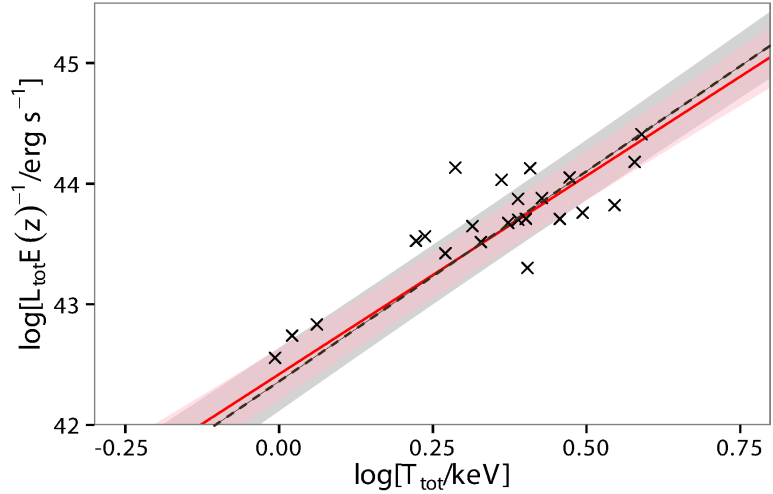}
 \caption[justified]{Bolometric LTR of 400d groups sample with bias-correction. The bias corrected model is represented by the red solid line, and the pink shaded region. 
The black dashed lines and the gray dashed region represents our result.    
  }
 \label{fig:comparison.4}
\end{figure*}

\section{An improved ``punctuated equilibria model"}\label{sec:pem}

In this section, we extend the ``punctuated equilibria model" (PEM) 
from 
\citep{Cavaliere1997,Cavaliere1998,Cavaliere1999}, taking into account the 
angular momentum acquisition from the proto-structure, and for the first time the 
DF
. 

In \cite{Cavaliere1998}, the cluster evolution is described as a sequence of ``punctuated equilibria" (PE). These are a sort of sequence of hierarchical merging episodes of the DM halos, related in the intra-cluster plasma (ICM) to shocks of various strengths
providing the boundary conditions so that the ICM may re-adjust to a new hydrostatic
equilibrium.

Following \cite{Balogh1999}'s notation, one can obtain the bolometric X-ray luminosity from Bremsstrahlung. If the cluster gas temperature is $T_\Mov{g}(r)$, its density profile $\rho_{\rm g}(r)$, we can write
\begin{equation}
L ={6 \pi k \over C_1(\mu m_{\rm p})^2}
\int_0^{R_{\rm vir}}r^2\rho_g(r)^2T_g(r)^{1/2}dr
\label{eq:bol}
\end{equation}
\cite{Balogh1999}, where $\mu=0.59$, $C_1=3.88 \times 10^{11}$s K$^{-1/2}$ cm$^{-3}$.
%
Assuming that 
$\rho_{\rm g} \propto \rho$ one obtains the simplest model describing the L-T relation that can be calculated by Eq.~(\ref{eq:bol}), namely the self-similar-model of \cite{Kaiser1986}.

In \cite{Cavaliere1998}'s notation, Eq.~(\ref{eq:bol}), is written \Mov{with }
a similar form 
\begin{equation}
L\propto \int_o^{r_2}\,n^2(r)\,T^{1/2}(r)\,d^3r~.
\end{equation}

Here $T(r)$ is the temperature in the plasma. A cluster boundary, $r_2$, is considered that is taken to be close to the virial radius $R_{\rm vir} \propto M_{\rm vir}^{1/3}\,\rho^{-1/3}$.
%

As shown 
in App.~\ref{sec:appendixA}
, the luminosity-mass relation
can be cast in the form:
\begin{align}
L \propto & \Big({n_2\over n_1}\Big)^2\,
\rho\,
\Bigg[{T_2\over T_v}\Bigg]^{1/2}\,
\overline{[n(r)/n_2\big]^{2+(\gamma-1)/2}}
M^{4/3} \nonumber\\
 & 
\sqrt {
\frac{
\frac{1}{m}+\left ({\frac {t_{{\Omega}}}{t}}
\right )^{2/3}+
\frac{K}{
\left ({\frac {M}{M_{{0}}}}\right )^{8/3}
}+\frac{\lambda_0}{1-\mu(\delta)} 
\frac{\Lambda \xi^3}{3 H_0^2 \Omega_m,0}
}{
\frac{1}{m}+\left ({\frac {t_{{\Omega}}}{t}}\right )^{2/3}+{\frac
{K_{{0}}}{{M_{{0}}}^{8/3}}}}
}
\label{eq:lll}\Mov{.}
\end{align}
See App.~\ref{sec:appendixA} 
for a derivation of
Eq.~(\ref{eq:lll}) and a definition
of the terms involved.
From the luminosity-mass relation the LTR can be obtained using the relation between mass and temperature \citep[see also][]{DelPopolo2005}.

We may now calculate the average value of $L$ and its dispersion,
associated with a given cluster mass.

To this aim, we sum over the shocks produced at a time $t'<t$
in all possible progenitors $M'$ 
by the accreted clumps $\Delta M$; 
then, we integrate over time $t'$ from an effective lower limit
$t-\Delta t$.
As found by \cite{Cavaliere1999} the average $L$ is given by
\begin{eqnarray}
\langle L\rangle&= & Q\,\int_{t-\Delta t}^{t}\,dt'\,\int_{0}^{m}\,dM'\,\int_{0}^{M-M'}\,d\Delta M\,\nonumber \\
 && \frac{df}{dM'}(M',t'|M,t)\,\frac{d^{2}p(M'\rightarrow M'+\Delta M)}{d\Delta M\,dt'}\,L~;\nonumber \\\label{eq:lum}
\end{eqnarray}
and the variance is given by
\begin{eqnarray}
\langle \Delta L^2\rangle &=& Q\,\int_{t-\Delta t}^t\,dt'
\int_0^M\,dM'\,\int_0^{M-M'}\,d\Delta M\, \nonumber\\
& &
\,{df\over dM'}(M',t'|M,t)\,
{d^2 p(M'\rightarrow M'+\Delta M)\over d\Delta M\,dt'} \nonumber\\
& &
\Big(L-\langle L\rangle\Big)^2~.
\label{eq:lumm}
\end{eqnarray}
where $Q$ is the normalisation factor \footnote{The compounded probability distribution in Eqs. (\ref{eq:lum}) and (\ref{eq:lumm}) has been normalised to 1.}.

The effective lower limit for the integration over masses is set
as described in \cite[Sec.~(2.4)]{Cavaliere1999}.

\section{Results and discussion}\label{sect:results}

The results of the calculations are represented in Figs.~\ref{fig:comparison.1}-\ref{fig:comparison.4}. In Fig.~\ref{fig:comparison.1} we plot the comparison 
with our present model 
of the LTR of \citep{Giles2016} represented by the dots with error-bars, and the gray line. The plot shows the bolometric LTR. The data are corrected for the bias, the solid line is the best fitting model, and the gray band represents the 1~$\sigma$ uncertainty. The red band represents our result, while the brown line is the prediction of the self-similar model, $L \propto T^2$. As shown by the plot, our model is in agreement with 
the survey 
of \citep{Giles2016}, but 
in contrast with their best fit model, 
it shows a slight break at $\simeq 2$ keV. The upper line has a slope of $2.76 \pm 0.18$, while the bottom one $3.4 \pm 0.18$. The \citep{Giles2016} bolometric LTR has a slope $3.08 \pm 0.15$. The slopes are in agreement 
with
in the limits of errors. 
%
%
As discussed previously, the break, and the different slopes at high mass and low mass 
are 
in agreement with the prediction of some observations \citep{Helsdon2000,Sun2009}. The same kind of break is observed in the MTR relation 
shown in Fig.~\ref{fig:comparison.2} \citep[see also][]{DelPopolo2019}. The MTR 
presents 
a non-self similar behavior with a break around 3 keV. At smaller masses the central slope of the MTR in the range $0.5-3$ keV is $ \simeq 2.3$. 
Such 
bend has been observed by \citep{Finoguenov2001}, who explained 
it 
as due to the formation redshift. Similarly to the LTR, another possibility to explain the bend in the MTR 
relies on the preheating of 
the ICM 
in the early phase of formation \citep{Xu2001}. As noticed by \citep{Giles2016} there is a dependence between the LTR and the MTR. The 
trends of one somehow are reflected in the other. 
As shown by the Fig.~\ref{fig:comparison.1}, both 
\citep{Giles2016}'s result, and our's show that the slope of the LTR is steeper than the self-similar case, $L \propto T^2$. They tend to 
appear closer 
to $L \propto T^3$. We have seen that this large steepness is due to different possible reasons
, such as 
pre-heating, supernovae feedback, or outflows generated by AGN at high redshift that blows cavities, and AGN heating, raising the entropy of the ICM, 
thus 
reducing its density, and consequently the X-ray luminosity, 
by that 
removing the gas in a large area 
reaching 
the cluster virial radius
. 
In other words, for some physical reasons, the core gas density is smaller than what is expected in the self-similar model. 
This can be understood more easily by taking the example of 
an episode of supernovae feedback heating the gas 
that 
could make it harder to compress in the core. 
Preheating or similar models then 
give rise to a steeper LTR because they change the density in the center of the cluster. As discussed in \citep{DelPopolo2005}, the same effect can be produced by angular momentum acquired by the cluster. As discussed in \citep{DelPopolo1998}, angular momentum acquired by a shell of matter centered in a peak of the CDM density distribution is anti-correlated with density. Namely, high density peaks acquire less angular momentum than low-density peaks \citep{Hoffman1986a,Ryden1988}. Larger amounts of angular momentum acquired by low-density peaks 
lead these peaks to 
resist more easily \Mov{to }gravitational collapse. In our model, we added the effect of 
DF
, \Mov{compared }with 
\citep{DelPopolo2005}, and as shown, for example in \citep[App.~D, and Fig.~11]{DelPopolo2009}, 
DF 
amplifies the effects of angular momentum.  

A more extended discussion of the reason 
why the 
LTR relation is steeper than the self-similar relation, $L \propto T^2$, follows. 

In 
our model the breaking of 
self-similarity is due to tidal interactions between neighboring perturbations, the predecessor of the clusters. The interaction arises due to the asphericity of clusters, and to the effect of 
DF
. 
The 
relation between the acquisition of angular momentum, asphericity, and structure formation 
was discussed in \citep{DelPopolo1999}
. Apart 
from 
the steepening from $L \propto T^2$ to $L \propto T^{\simeq 3}$, as shown in Fig
s.~\ref{fig:comparison.1} and \ref{fig:comparison.2}, a bend around a few keV can be found. That 
bend depends \Mov{on }
the asphericity, \Mov{and appears }both in the LTR and in the MTR. Then, for clusters having $T< 2 \rm keV$, the luminosity is smaller than 
for clusters with higher 
energies. The difference with the self-similar solution increases going towards smaller masses. Our MTR, and LTR are characterized by a mass dependent angular momentum, $L$, that originates 
in 
the quadruple momentum of the proto-cluster. This is responsible of the breaking of 
self-similarity, 
to which 
we must add the effect of 
DF 
that behaves similarly to 
angular momentum, delaying the collapse of the cluster. 
In more details, 
the collapse 
in our model 
occurs in a different way 
than in the case of the simple spherical collapse model. In our model, the collapse threshold, $\delta_c$, and the time of collapse change. Differently from the simple spherical collapse, the threshold $\delta_c$ becomes mass dependent as shown in several papers [e.g., \citealt{DelPopolo1999}: Eq.~14, Fig.~1; \citealt{DelPopolo2006b}: Fig.~5], and becomes a monotonic decreasing function of mass, as shown in Fig.~\ref{fig:comparison.3} 
of the present paper. As the plot shows, $\delta_c$ is larger with respect to the standard value at the galactic mass scale, and moves to the standard value, going towards 
cluster mass 
scale
s. In Fig.~\ref{fig:comparison.3}, 
the red solid line shows the result \Mov{from }
\citep[Eq.~4]{Sheth2001}, the green short-dashed line, the results of \citep{DelPopolo1998} taking into account the effect of the tidal field, the orange long-dashed 
line, 
the result of \citep{DelPopolo2006c} taking into account the effect of the tidal field and the cosmological constant, while the blue dot-dashed line takes into account the effect of the tidal field, the cosmological constant and 
DF. As can be seen from the plot, the effects of tidal torques, cosmological constant, and 
DF 
are additive, and tend to make $\delta_c$ increase.  

Moreover there is a correlation between the temperature and threshold $\delta_c$: $T \propto \epsilon \propto \delta_c$ 
\citep[see][]{Voit2000}. This means that less massive clusters are hotter than more massive ones, characterized by a standard MTR.

Another factor 
modifying the LTR, as well as the MTR, is the modification of energy partition in virial equilibrium \citep{DelPopolo1999}. Another fundamental role is taken 
by 
the cosmological constant, and 
DF
. The effect of these two quantities 
are also similar to that of 
angular momentum
, namely that of delaying the collapse of the perturbation, and modifying $\delta_c$. 
A comparison of angular momentum, cosmological constant, and 
DF
, represented by the three terms after the gravitational 
force 
in Eq.~(\ref{eq:coll}), is shown in Fig.~\ref{fig:comparison.3}. These quantities have a similar order of magnitude, with 
differences of order of 
a few percent. The effect of 
DF 
is calculated as in \citep{DelPopolo2006b,DelPopolo2009}. There are several papers studying the role of 
DF 
in clusters formation \citep{White1976,Kashlinsky1984,Kashlinsky1986,Kashlinsky1987}. The effect of substructure was calculated by \citep{AntonuccioDelogu1994}, who also showed that 
DF 
gives rise to a delay in the collapse of low height ($\nu$) peaks, with consequences similar to that of tidal torques. 
DF 
and tidal torques \Mov{produce }
changes in several quantities like: the threshold of collapse $\delta_c$, the mass function, the temperature at a given mass, and the correlation function. 
DF 
has a similar effect on structure formation to that of tidal torques. It slows down the collapse, and as a consequence the density in the central 
region of \Mov{the halo }
decreases.

We have seen that both the MTR and the LTR at some keV show a steepening, and 
that clusters with lower mass than the bend 
are hotter (less luminous) than expected. To further check the LTR at the low mass regime, we compared the results of our model with that of a study by \citep{Zou2016}, dealing with $Chandra$ observations of 23 galaxy groups and low-mass clusters of galaxies. 
The comparison of our results with 
the clusters from 
\citep{Zou2016} 
is 
plotted in Fig.~\ref{fig:comparison.4}. The figure plots the bolometric LTR of the 400d groups sample with bias correction, 
as 
the red solid line with the pink shaded error region. The black dashed line and the gray dashed region represent our results. In 
\citep{Zou2016}, 
the LTR relation 
was studied
, finding that 
its 
slope 
is more or less in agreement with that of large mass clusters, 
evaluated at 
$3.29 \pm 0.33$. In our case we 
obtained 
$3.35 \pm 0.3$. From a comparison with the clusters of the XXL survey of \citep[$3.08 \pm 0.15$]{Giles2016}, it is evident that the maximum slope is in this last case 3.23, while for the case of the \citep{Zou2016} groups it reaches 3.62. In other terms, even if the groups, and clusters have similar behavior, groups seems to have a steeper slope in agreement with other studies \citep{Helsdon2000,Sun2009}, and in agreement with our results. 

\Mov{
Note that the X-ray luminosity, at energies close to 1 KeV, especially at the level of groups where the emission in the Fe-L band becomes very relevant, depends on the assumed metallicity \cite{Giles2016,Zou2016,Sun2009,Eckmiller:2011gv,Maughan2012}:
the larger the metallicity, the larger $L_X$ is. In this paper we have employed the solar metallicity (0.3) from emission models such as APEC \cite{Smith:2001he} or XSPEC \cite{Arnaud1996XSPEC}.}

\section{Conclusions}\label{sect:conclusions}

In the present work, we obtained 
an 
LTR extending the results of a previous paper \citep{DelPopolo2005}, 
additionally 
taking into account in the model the effects of 
DF
. 
Under the assumption of self-similarity \citep{Kaiser1986}, one can obtain a simple 
scaling 
law between the luminosity and temperature. The relation is of the type $L \propto T^2$. A large number of studies have shown that in reality the relation between luminosity and temperature is steeper, $L \propto T^ {\simeq 3}$. There are several reason 
for 
this discrepancy between theory and observations. The difference is mainly due to energy input by non-gravitational processes, like pre-heating, supernovae feedback, and heating from AGN. In this paper, we studied the LTR by means of a modified version of the \citep{Cavaliere1999} punctuated equilibria model, taking
, in addition, DF 
into account
, 
while it was ignored 
in \citep{DelPopolo2005}. We showed that the model is predicting a non-self-similar behavior of the LTR. The LTR 
features 
a bend at $\simeq 2$ keV, with a slope $2.76 \pm 0.18$ 
for 
larger energies and $3.4 \pm 0.18$ 
for 
energies smaller than 2 keV. This behavior is in agreement with the result of the XXL survey \citep{Giles2016}, 
shifting the slope to 
$3.08 \pm 0.15$. The steeper slopes at smaller energies is in agreement with some studies  \citep[e.g.]{Helsdon2000,Sun2009} claiming a further steepening of the LTR at the low mass end. In order to check 
such slightly steeper LTR slope 
in small mass clusters and groups of galaxies
, we used the study of 
\citep{Zou2016} for the 400d groups sample, and compared our result with them. This study, in agreement with ours, 
indeed displays 
a slope $3.29 \pm 0.33$,\footnote{In our case $3.35 \pm 0.3$} 
steeper than clusters of the XXL survey 
from 
\citep[$3.08 \pm 0.15$]{Giles2016}.

\section*{Acknowledgements}

MLeD acknowledges the financial support by the Lanzhou University starting
fund, the Fundamental Research Funds for the Central Universities
(Grant No. lzujbky-2019-25), National Science Foundation of China (grant No. 12047501) and the 111 Project under Grant No. B20063.
\appendix
\section{
L-T relation in the MPEM}\label{sec:appendixA}

As we saw in 
Sec.~\ref{sec:pem} that Eq.~(\ref{eq:bol}) 
gives the X-ray bolometric luminosity of a cluster. Using \cite{Cavaliere1998}'s notation the equation is written as
\begin{equation}
L\propto \int_o^{r_2}\,n^2(r)\,T^{1/2}(r)\,d^3r~.
\end{equation}
$T(r)$ indicates the plasma temperature. The cluster has a boundary $r_2$ close to the 
virial radius $R_{\rm vir} \propto M^{1/3}\,\rho^{-1/3}$
%
%
and at this radius the gas is supersonic  forming a shock front \cite{Tazikawa1998}.
Mass, and energy conservation is represented by the Rankine-Hugoniot conditions, describing the jumps in temperature and density from the outer region of the shock where temperature and density are $T_1$ and $n_1$ to $T_2$ and $n_2$ in the region interior to $r_2$.
Then we may write the luminosity as 
\begin{equation}
L\propto r_2^3\;n_2^2\,T_2^{1/2}\;
\int_0^1 d^3x\,\Big[{n(x)\over n_2}\Big]^2\,
\Big[{T(x)\over T_2}\Big]^{1/2}~,
\end{equation}
 where $x\equiv r/r_2$.
$n_1$ is proportional to the universal baryonic fraction $f_u$, and 
average cosmic DM density $\rho_u$ at formation, taking the form $n_1 \propto f_u \, \rho_u /m_p$. The inner temperature $T_1$ is obtained in a statistical way by means of 
the diverse merging histories ending up in the mass $M$. 
In summary, for each dark mass $M$ exists a set of equilibria states of the intra-cluster plasma (ICP) corresponding to different initial conditions. Each one of these equilibria states correspond to a  different realization of the merging history. The average values of luminosity, $L$, and $R_X$, and their scatter are obtained from the convolution over 
such 
set. 
According to \cite{Cavaliere1998}, in a merging event the temperature before the shock, during a merging, is considered equal to that of the infalling gas. If the latter is inside a deep potential well, $T_1$ is the virial temperature $T_{1v}\propto \Delta M/r$
of the partner in secondary merging. If $r\propto (\Delta M/\rho)^{1/3}$, the virial temperature fulfills the relation 
\begin{equation}
k\,T_{1v}=4.5\,(\Delta M)^{2/3}\,(\rho/\rho_o)^{1/3}~{\rm keV},
\end{equation}
where the masses 
$M/M_0$ are normalised to the mass inside a sphere of $8 h^{-1}$ Mpc, giving the current value $M_0=0.6 \times 10^{15} \Omega_0 h^{-1} M_{\odot}$, and the numerical factor comes from \cite{Hjorth1998}. 

Preheating of diffuse external gas, generated by feedback energy inputs (e.g., supernovae) gives rise to a lower bound $kT_{1*}\approx 0.5$ keV \cite{Renzini1997}.

Then the actual value of $T_1$ is 
\begin{equation}
T_1=max\,\big[T_{1v},T_{1*}\big] 
\end{equation} 

Given the $T_1$ value, the boundary conditions for the ICP is given by the shocks strength. 
The value of $T_2$ in the case of a nearly hydrostatic post-shock condition with $v_2<< v_1$, three degrees of freedom, and 
the velocity of the shock 
that 
matches the growth rate of the virial radius, 
follows
\begin{equation}
kT_2={{\mu m_p v_1^2}\over 3}\Big[ {{(1+\sqrt{1+\epsilon})^2}\over 4}
+ {7\over{10}}\epsilon -{{3}\over {20}}{{\epsilon^2}\over{(1+\sqrt{1+
\epsilon})^2}}\Big]\, .
\label{eq:kkt}
\end{equation}
\cite{Cavaliere1997}, where $\epsilon\equiv 15 kT_1/4 \mu m_p v_1^2$ and $\mu$ is the average
molecular weight. $v_1$, namely the inflow velocity is fixed by the 
potential drop across the region of free fall, and is given by 
$v_1 \simeq \sqrt{-\phi_2/m_p}$, being $\phi_2$ the potential at $r_2$.
For strong shocks (``cold inflow") $\epsilon <<1$. 
The approximation
 \begin{equation}
k T_2 \simeq -\phi_2/3 +3 k T_1/2
\label{eq:kt}
\end{equation}
holds in the case of rich clusters accreting diffuse gas in small clumps. Here, 
$\phi_2$ represents the gravitational potential energy at $r_2 \simeq R_{\rm vir}$.
In the case of weak shocks ($\epsilon \geq 1$), we have that $T_2 \simeq T_1$. 
Following \cite{Cavaliere1997}, the density jump is given by 
\begin{equation}
{n_2\over n_1} =
2\,\Big(1-{T_1\over T_2}\Big)+\Big[4\,
\Big(1-{T_1\over T_2}\Big)^2 + {T_1\over T_2}\Big]^{1/2}~.
\end{equation}\vspace{.000cm}

With the polytropic temperature description $T(x)/T_2= [n(x)/n_2]^{\gamma-1}$, where $1\leq \gamma\leq 5/3$, and assuming that $r_2$ may be related 
to it, 
written as $T_{\rm v} \propto M/r_2$
and that $M \propto \rho r_2^3$, (leading to $r_2 \propto (t/\rho)^{1/2}$), we get the luminosity in the form 
\begin{equation}
L\propto \Big({n_2\over n_1}\Big)^2\, M T_v^{1/2}\,\rho\,
\Bigg[{T_2\over T_v}\Bigg]^{1/2}\,
\overline{[n(r)/n_2\big]^{2+(\gamma-1)/2}} ~.
\label{eq:ll}
\end{equation}
The bar in the previous expression represents the integration over the emitting volume $r^3\leq r_2^3$, while $\rho$ (proportional to $\rho_u$ and so to $n_1$)
is the average DM density in the cluster. 

Eq.~(\ref{eq:ll}) can 
also 
be 
cast in the form:

\begin{eqnarray}
L &\propto & \Big({n_2\over n_1}\Big)^2\,
\rho\,
\Bigg[{T_2\over T_v}\Bigg]^{1/2}\,
\overline{[n(r)/n_2\big]^{2+(\gamma-1)/2}}
M^{4/3} \nonumber\\
& & 
\sqrt {
\frac{
\frac{1}{m}+\left ({\frac {t_{{\Omega}}}{t}}
\right )^{2/3}+
\frac{K}{
\left ({\frac {M}{M_{{0}}}}\right )^{8/3}
}+\frac{\lambda_0}{1-\mu(\delta)} \frac{\Lambda \xi^3}{3 H_0^2 \Omega_m,0}
}{
\frac{1}{m}+\left ({\frac {t_{{\Omega}}}{t}}\right )^{2/3}+{\frac
{K_{{0}}}{{M_{{0}}}^{8/3}}}}
}
\label{eq:llll}
\end{eqnarray}
where $\lambda_0=\epsilon_0 T_{\rm c0}$, $\mu(\delta)$, $\xi$, $M$, $t_\Omega$, and the function $K$ are given in 
Sec.~\ref{sect:model} 
of the present paper.

Using the hydrostatic equilibrium, $
dP/
m_p\,n\,
dr=-G\,M(<r)/r^2=-d\phi/dr$, together with the 
polytropic pressure $P(r)= kT_2\,n_2\,\big[{n(r)/n_2}\big]^{\gamma}$, we obtained the ratio $n(x)/n_2$. This gives rise to the profiles 
\begin{equation}
{n(r)\over n_2}=\Big[{T(r)\over T_2}\Big]^{1/(\gamma-1)}=
\Big\{1+{\gamma-1\over \gamma}\,\beta\,
\big[\tilde{\phi}_2-\tilde{\phi}(r)\big]\Big\}^{1/(\gamma-1)}~,
\label{eq:dens}
\end{equation}
being $\tilde{\phi}\equiv \phi/\mu\,m_p\,\sigma_2^2$ the potential 
normalised to the relative one-dimensional DM velocity dispersion at the radius $r_2$. The parameter $\beta$ is given by 
\begin{equation}
\beta = \mu m_p \sigma_2/kT_2~.
\label{eq:mu}
\end{equation}
%
%
Eq.~(\ref{eq:kt}) for a given 
DM 
potential $\phi_2$ allows to obtain the function 
$\beta(T)$ corresponding to $\rho(r)$, and $\phi(r)$
, while 
$\sigma(r)$ can be obtained by means of 
Navarro-Frenk-White profile \citep{Navarro1997}.
$\beta(T)$ goes from $\beta \simeq 0.5$ for $T \simeq T_1$ to $\beta \simeq 0.9$ for $T>> T_1$. Following \citep{Cavaliere1999}, and the observations of \cite{Markevitch1998} $\gamma =1.2 \pm 0.1$. We will use the value $\gamma=1.2$ in the calculation.

%
\bibliographystyle{apsrev4-1}
\bibliography{old_MasterBib}

\end{document}